# Charge carrier coupling to the soft phonon mode in a ferroelectric semiconductor

Mark E. Ziffer, Lucas Huber, Feifan Wang, Victoria A. Posey, Jake C. Russell, Taketo Handa, Xavier Roy, and X.-Y. Zhu[*]

Department of Chemistry, Columbia University, New York, NY 10027, USA

**Abstract**

Many crystalline solids possess strongly anharmonic "soft" phonon modes characterized by diminishing frequency as temperature approaches a critical point associated with a symmetry breaking phase transition. While electron-soft phonon coupling can introduce unique scattering channels for charge carriers in ferroelectrics, recent studies on the non-ferroelectric lead halide perovskites have also suggested the central role of anharmonic phonons bearing resemblance to soft modes in charge carrier screening. Here we apply coherent phonon spectroscopy to directly study electron coupling to the soft transverse optical (TO) phonon mode in a ferroelectric semiconductor SbSI. Photo-generated charge carriers in SbSI are found to be exceptionally long lived and are associated with a transient electro-optical effect that can be explained by interactions between charge carriers and thermally stimulated soft-phonon excitations. These results provide strong evidence for the role of electron-soft phonon coupling in the efficient screening of charge carriers and in reducing charge recombination rates, both desirable properties for optoelectronics.

**Introduction**

Recent studies have suggested that the interplay between lattice and charge carrier dynamics in strongly anharmonic polar semiconductors, such as lead halide perovskites, could explain their outstanding optoelectronic properties.[1–6] These findings have recently led to a "ferroelectric polaron" proposal,[7,8] in which electron coupling to polar anharmonic phonon modes bearing resemblance to soft-phonons in ferroelectrics may lead to local ordering of dipoles from unit cells with broken symmetry.[7,8] Generally speaking, soft phonons refer to a lattice mode associated with a symmetry-breaking atomic displacement, and exhibit a temperature dependent vibrational frequency diminishing towards zero as a symmetry-breaking phase transition is approached.[9–11] Soft phonons are particularly important in displacive-type ferroelectric crystals, wherein the ferroelectric-paraelectric phase transition is driven by the softening of a phonon mode whose normal coordinate corresponds to atomic displacement along the symmetry-breaking axis of spontaneous polarization.[9,12,13] Historically, electron-soft phonon coupling in ferroelectrics has been studied with regards to phase transitions, where temperature dependent vibrational mixing between electronic bands mediated by a soft phonon mode has been proposed to drive the ferroelectric-paraelectric phase transformation.[14,15] Additionally, electron-soft phonon coupling in ferroelectrics has been studied in wide-band gap ferroelectric crystals such as perovskite oxides to understand the scattering interactions that influence charge carrier mobility and electrical transport.[16–20]

---

[*] To whom correspondence should be addressed: xz2324@columbia.edu

For electron-phonon coupling in conventional semiconductors, one is commonly concerned with the long range Fröhlich interaction, which describes the Coulomb interaction between an electron and the macroscopic electric field in the lattice associated with atomic displacement along a polar longitudinal (LO) phonon mode.[21,22] However, the soft phonon in ferroelectrics is a TO-phonon mode according to the Lyddane-Sachs-Teller relation[10,12] and consequently does not give rise to a long range electric field, preventing coupling with electrons via the Fröhlich interaction.[8,16] Instead, electron coupling with soft TO-phonons has been addressed by Wemple *et al.* as involving the scattering of charge carriers via the polarization potential interaction,[16,23,24] by Nasu and co-workers in the formation of small and large "superparaelectric" polarons,[25,26] and by Bernardi and co-workers in the direct charge carrier scattering studied with ab-initio dynamics.[17,18]

Most experimental studies on electron-soft phonon coupling in ferroelectric crystals have relied on analyzing temperature dependent charge transport phenomena.[16–20] These measurements do not directly probe mode-specific electron-phonon interactions, and the involvement of soft phonons often comes from qualitative comparisons to theoretically predicted temperature dependences in mobility. Furthermore, few experiments have directly studied the role of electron-soft phonon coupling in the photophysics of semiconductors, and ferroelectric semiconductors provide an excellent model system to explore such phenomena. Here, we study mode-specific electron-phonon coupling in a ferroelectric semiconductor, SbSI, using ultrafast pump-probe coherent phonon spectroscopy.[27–29] Using a broadband probe with photon energies across the band gap of SbSI, we directly identify coupling between photoexcited electrons and coherently excited soft phonons. Furthermore, we find evidence from transient reflectance spectroscopy to suggest that charge carrier dynamics in SbSI are strongly influenced by coupling between free charges and thermally stimulated soft-phonon excitations, resulting in long charge carrier recombination lifetimes associated with a transient electro-optical effect.

**Results and Discussion**

*Soft-mode characterization and electron soft-phonon coupling.* Single crystal SbSI is a prototypical ferroelectric semiconductor that features a displacive-type phase transition from a low symmetry $C_{2v}^9$ (*Pna*2$_1$) ferroelectric phase to a high symmetry $D_{2h}^{16}$ (*Pnam*) paraelectric phase at ~290 K, driven by a ferroelectric soft TO-phonon mode.[10,30–32] Figure 1a shows the crystal structure of SbSI in the ferroelectric phase[33] with several unit cells stacked vertically along the crystallographic *c*-axis, which corresponds to the axis of spontaneous polarization.[34] Previous studies have identified the soft mode in SbSI as an IR and Raman active TO-phonon involving the displacement of Sb-S atoms along the *c*-axis.[32] Polarization resolved Raman scattering measurements and group theory calculations have shown that the soft phonon mode is of $A_u$ symmetry, with a Raman tensor consisting of an α$_{zz}$ component, where *z* is the coordinate along the *c*-axis.[30,32] We use polarization resolved Raman spectroscopy to first characterize the soft-phonon mode in flux-synthesized single crystal SbSI (see Methods). Figure 1b shows the polarization resolved Raman spectrum of SbSI measured at 80 K in the backscattering geometry, with incident polarization in the *xz* plane and signal collected for scattered polarization parallel to the incident polarization. Here *x* is in the direction of either the *a*- or *b*-axis, which we do not distinguish since the Raman responses for polarization incident in the *c-a* and *c-b* planes in SbSI are known to be similar.[35] Figures 1c and 1d show polar plots for the polarization dependent Raman signal taken from cuts along the two dominant modes at ~50 cm$^{-1}$ and ~115 cm$^{-1}$ in Figure 1b. Both the frequency and the symmetry for the parallel polarized Raman scattering of the 50 cm$^{-1}$ mode at 80 K are in good agreement with those of the ferroelectric soft mode identified in previous experiments.[30,32] The polarization dependence of the 115 cm$^{-1}$ mode, which does not soften with temperature, is also in good agreement with previous studies.[30] Figure 1e confirms the characterization of phonon softening for the 50 cm$^{-1}$ mode identified at 80 K, shown by the decrease in vibrational frequency when the sample temperature is increased towards the ferroelectric phase transition temperature at 290 K.[30] Furthermore, the broadening of the soft-mode spectrum with temperature and appearance of a central-peak towards zero wavenumber at 290 K (Figure 1e) are clear indications of temperature dependent anharmonic phonon-phonon interactions that dampen the vibrational resonance and drive phonon softening in displacive ferroelectric crystals.[10,36]

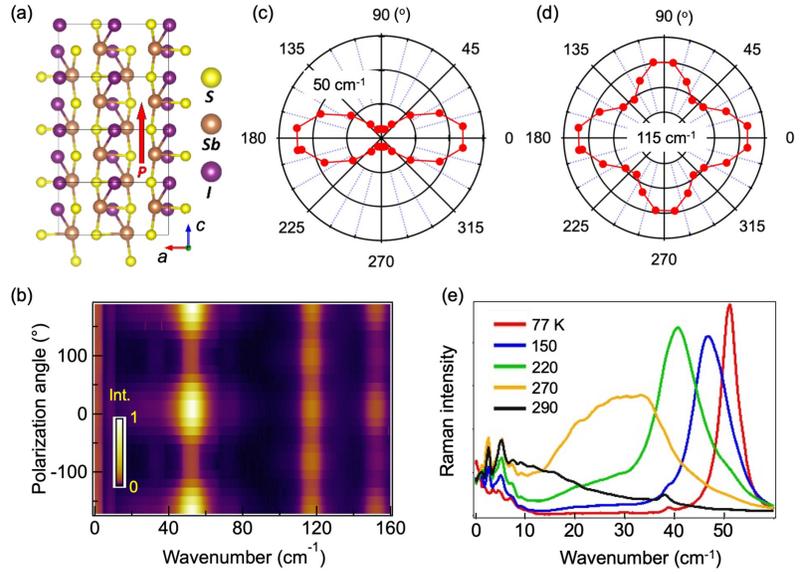

**Figure 1. Raman spectra identifying the soft phonon mode in SbSI.** (***a.***) Crystal structure of SbSI in the ferroelectric $C_{2v}^9$ (*Pna*2$_1$) phase. The spontaneous polarization is directed along the crystallographic *c*-axis as indicated by the red arrow. (***b.***) Polarization-angle resolved Raman spectrum of SbSI at 80 K with incident polarization in the *xz* plane (*z*=*c*-axis, *x* =*a* or *b*-axis) and scattered polarization parallel to incident polarization. Color bar indicates the Raman scattering intensity (a.u.) (***c. & d.***) Polar plots corresponding to spectral cuts from ***b.*** integrated around the ~50 cm$^{-1}$ and ~115 cm$^{-1}$ modes. (***e.***) Temperature dependence of the parallel polarized Raman scattering showing softening of the 50 cm$^{-1}$ (80 K) phonon mode.

To study electron-phonon coupling with the soft-mode, we use ultrafast pump-probe coherent phonon spectroscopy. We excite SbSI above the bandgap using a ~200 fs laser pulse centered at $h\nu_1$ = 3.1 eV (15 µJ/cm$^2$), with pump polarization parallel to the *c*-axis. The pump bandwidth is sufficiently broad to stimulate a wavepacket of coherent phonons with frequencies up to ~165 cm$^{-1}$ via the nonlinear impulsive stimulated Raman mechanism.[37–39] Figures 2a and 2b show the near-normal incidence transient reflectivity spectrum ($\Delta R/R$) of SbSI at 80 K probed by a ~100 fs broadband probe pulse ($h\nu_2$ ~1.75-2.5 eV) polarized either parallel or perpendicular to the crystallographic *c*-axis. The pump-induced phonon wave-packet modulates the complex optical susceptibility of SbSI, which results in coherent oscillations in the reflectivity ($\Delta R/R$).[37] To confirm that the coherent phonons come from the same modes characterized in Raman scattering, we analyze the below-gap coherent phonon response, which corresponds to modulation of the real part of the optical response due to the Raman susceptibility tensor.[37] Figure 2c shows spectral cuts of the pump-probe data from Figures 2a and 2b around time zero, showing derivative-like features from $h\nu_2 \sim$ 2.1-2.5 eV that can be attributed to an electronic response from SbSI due to the above-gap pump energy.[34,40] At $h\nu_2$ < 2.0 eV, the transient reflectance spectrum stays flat at $\Delta R/R \sim 0$, which is consistent with low temperature absorption measurements on SbSI showing that the lowest energy indirect absorption edge lies above 2.0 eV at 80 K.[41] We therefore assign the spectral range at $h\nu_2 \leq$ 1.9 eV to signal dominated by the below-gap coherent phonon response.

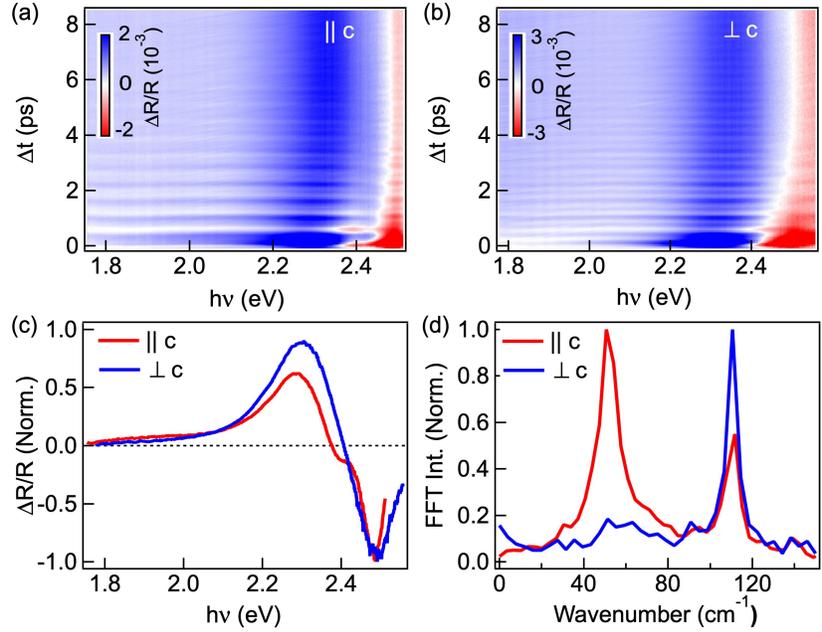

**Figure 2. Coherent phonon spectra of SbSI.** (**a. & b.**) Transient reflectivity of SbSI measured at 80 K for probe polarizations parallel and perpendicular to the *c*-axis, respectively. Color bars indicate the magnitude of $\Delta R/R$. (**c.**) Normalized spectral cuts of the transient reflectivity data in (**a.**) and (**b.**) integrated between $\Delta t$ =0 and 0.1 ps. (**d.**) Normalized FFT spectra of the time differentiated transient reflectivity data in (**a.**) and (**b.**), taken from signal transients spectrally integrated in the below-gap region (1.7-1.9 eV).

In Figure 2d we show the below-gap coherent phonon spectrum for probe-$\|c$ and -$\perp c$ polarization geometries, obtained by taking the FFT of the time differentiated transient reflectance signal ($\frac{d}{dt}\Delta R/R$) from Figures 2a and 2b and spectrally integrating for $h\nu_2 \leq$ 1.9 eV. A coherent phonon at 50±2 cm$^{-1}$ modulates the optical response only for probe polarization parallel to the *c*-axis, which is consistent with the symmetry of the Raman tensor for the ferroelectric soft mode identified from Raman scattering (Figure 1c). Additionally, we note that the coherent phonon at 115±2 cm$^{-1}$ is detected in both probe-∥$c$ and probe-⊥$c$ polarization geometries, also in agreement with the polarization dependence of the Raman scattering data (Figure 1d). Furthermore, we find that the dephasing lifetime of coherent oscillations from the soft mode in the probe-∥$c$ polarization decreases as the sample temperature increases from 80 K to 270 K (Supporting Information Section S1), consistent with softening of this phonon mode as the ferroelectric phase transition is approached.[10] In particular, we find that at 270 K coherent oscillations from the soft phonon mode dephase on a time scale around ~1 ps (Supporting Information Figure S1c), corresponding to a resonance linewidth of ~30 cm$^{-1}$ that is in good agreement with the linewidth of the soft-mode peak observed in Raman scattering at 270K (Figure 1e). Both the polarization and temperature dependent analysis of the below-gap coherent phonon signal thus provide conclusive evidence that the impulsively stimulated coherent 50 cm$^{-1}$ (80 K) mode corresponds to the ferroelectric soft mode.

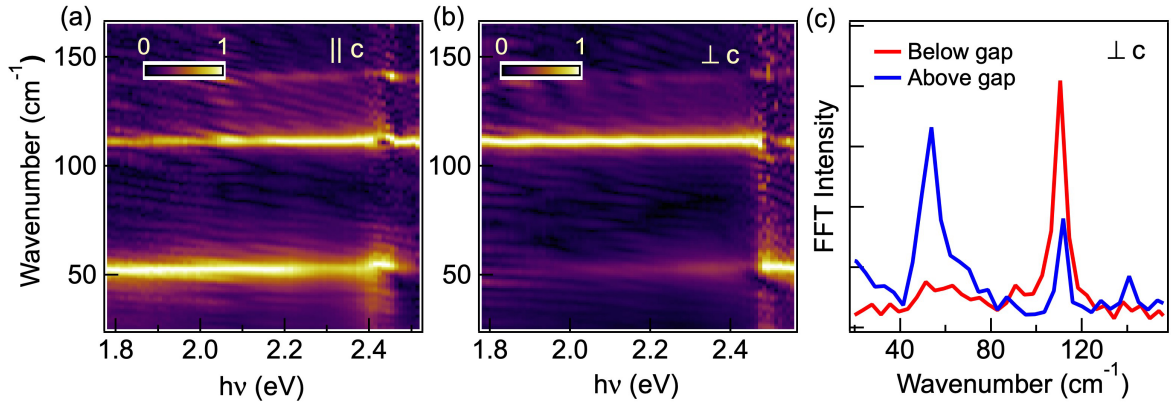

**Figure 3. Evidence for electron-soft phonon coupling in the above-gap probe of coherent phonons.** (**a. & b.**) Probe hv-resolved FFT of the coherent phonon signal for probe polarizations parallel (**a.**) and perpendicular (**b.**) to the $c$-axis. FFT spectra are amplitude normalized at each probe frequency and color bars indicate the normalized FFT intensity. (**c.**) Comparison of the FFT of the coherent phonon signal for probe frequencies below and above the bandgap in the probe-⊥$c$ polarization geometry. The below gap FFT is from the coherent phonon signal spectrally integrated between 1.7-1.9 eV and the above gap FFT is from signal spectrally integrated from 2.49-2.56 eV.

We next turn to analyze the coherent phonon response for probe frequencies above the band gap. Figures 3a and 3b show the hv$_2$ resolved FFT spectra of the coherent phonon signal from Figures 2a and 2b for probe polarizations parallel and perpendicular to the $c$-axis, respectively, with the FFT amplitude normalized at each probe energy. In the probe-∥$c$ polarization geometry (Figure 3a), both the 50 cm$^{-1}$ and the 115 cm$^{-1}$ modes are observed for hv$_2$ below and above the bandgap, consistent with the polarization dependence of the Raman susceptibility for these modes. However, in the probe-⊥$c$ geometry (Figure 3b), while the 50 cm$^{-1}$ soft mode is nearly absent at h$\nu_2$ below gap, it is switched on to become dominant above gap at h$\nu_2 \geq 2.47$ eV (Figure 3b). In Figure 3c, we highlight the difference between the above-gap and below-gap responses for the probe-⊥$c$ geometry by comparing the FFT spectra integrated in the 1.7-1.9 eV range with that in the 2.49 eV-2.56 eV range. The switching-on of coherent oscillations from the soft mode for h$\nu_2$ above the bandgap in the Raman symmetry forbidden $E \perp c$ probe polarization geometry can be attributed to the modulation of the optical susceptibility by electron-phonon coupling.[28] Atomic displacement along a phonon normal coordinate results in a change in the single particle band energies,[28] causing a strong modulation of the optical transitions that can be observed in the above-gap optical response in coherent phonon spectroscopy.[28,42] In ferroelectrics, such a change in the band energies due to atomic displacement along the soft-phonon mode coordinate is described by the polarization potential interaction,[16,34] and this interaction is particularly strong for bands at the $U$ point in SbSI that constitute the lowest energy direct optical transitions in the $E \perp c$ geometry probed in Figures 3a-c.[34,43] The modulation of the optical susceptibility due to the soft-mode in the above-gap coherent phonon data for $E \perp c$ (Figures 3b and c) therefore provides direct evidence for electron-soft phonon coupling in SbSI.

*Charge carrier recombination.* We now study charge carrier dynamics using the same pump-probe scheme as in Figures 2 and 3, but on longer time scales ($\Delta t \leq 3$ ns, limited experimentally by the delay stage). Figure 4a shows the transient reflectance data at 270 K in the probe-⊥$c$ polarization geometry. On the nanosecond time scale, transient reflectance signal above the bandgap is typically dominated by changes in the refractive index ($\Delta n$) due to the electronic response, which can be approximately converted to a change in extinction coefficient ($\Delta k$) via a Kramers-Kronig (K-K) transformation (Figure 4b).[40] The K-K transformed transient reflectance data shows a long-lived Gaussian-like peak centered around 2.3 eV (Figure 4b), corresponding to the lowest energy $U$-point direct-gap transition in SbSI.[34] Figure 4c plots the amplitude of this Gaussian peak in the K-K transformed signal as a function of $\Delta t$ at two excitation pulse energy densities, $\rho = 5$ and 18 µJ/cm$^2$. At a low excitation density of $\rho = 5$ µJ/cm$^2$, there is no measurable decay, indicating that the electronic process associated with this feature has a characteristic lifetime more than one order of magnitude longer than the 3 ns experimental limit. Only at the much higher excitation density of ρ = 18 μJ/cm² do we observe a measurable decay on the sub-nanosecond time scale, which could be attributed to many-body effects such as Auger recombination of photo carriers.[44] We note that the Gaussian feature in the K-K transformed data at 2.3 eV lies well above the indirect gap of SbSI at ~2.0 eV at 80K,[41] suggesting that the transient reflectance spectrum does not simply correspond to a ground state bleach of the lowest energy optical transition.

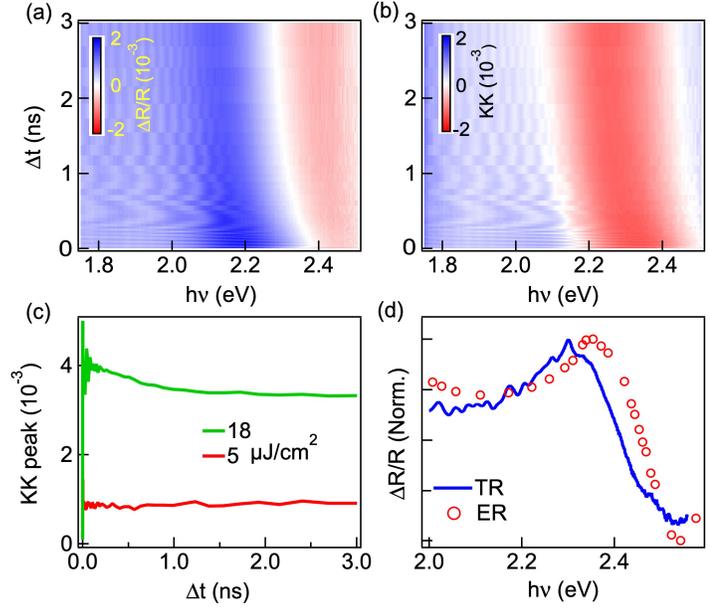

**Figure 4. Dynamics of charge carriers from transient reflectance.** (*a.*) Transient reflectance (*ΔR/R*, color scale) of SbSI at 270 K with probe polarization perpendicular to the *c*-axis. (*b.*) Kramers-Kronig (K-K) transform of the data in (*a.*), which gives an approximation to change in the extinction coefficient Δ*k* (color scale). (*c.*) Peak of the K-K transformed signal as a function of Δt at excitation pulse energies of ρ =5 μJ/cm² (red) 18 μJ/cm² (green). (*d*). Comparison between the transient reflectance spectrum measured at 80K (integrated for Δt = 0.5-3ns for, ρ = 15 μJ/cm², see Figure S2) for probe-⊥*c* and the electro-reflectance spectrum of SbSI for probe-⊥*c* at 90K reported previously.[45]

Interestingly, Figure 4d shows good agreement between the transient reflectance spectrum measured at 80 K with the electro-reflectance spectrum of SbSI measured by Zametin *et al.*[45] at 90 K in the same polarization geometry. The observation of an electro-optical response associated with photoexcited charge carrier dynamics in SbSI can be explained based on the model proposed by Wemple *et al.* for electron-soft phonon interactions.[16] Here, thermally stimulated incoherent excitations of soft-phonons with wavevector $q$ and amplitude $a_q$ give rise to fluctuating spatial regions of lattice polarization ($\delta P$), with a spatial correlation length given by:[16,46]

$$\langle \delta P(0) \delta P^*(r) \rangle \propto \sum_q \langle |a_q|^2 \rangle e^{iqr} \tag{1}$$

$$\langle |a_q|^2 \rangle \propto \frac{kT}{\omega_q^2} \tag{2}$$

The finite spatial correlation length (Eq. 1) of these "polarization clusters" gives rise to nonzero polarization gradients at their boundaries, resulting in the presence of surface bound polarization charge ($\nabla \delta P = \sigma_b$) and corresponding cluster depolarization fields ($\delta E_d$).[12,16] Following above gap photoexcitation, free charge carriers neutralize the surface bound charge which results in a screening of $\delta E_d$ and a subsequent change in $\delta P$ of the polarization cluster.[16,47] A change in $\delta P$ causes a modulation of the band energies near the low energy direct gap, due to the large polarization potential of bands at the $U$ point in SbSI. This interaction is analogous to the electro-optical effect in ferroelectrics,[23,34,45] and results in the shift in optical transition energies at the direct gap $U$ point in SbSI that is seen in both the electro-reflectance and transient reflectance spectroscopy (Figure 4d). We note that dynamic changes in lattice polarization due to the screening of bound polarization charge by photoexcited free carriers also have been observed directly in ferroelectric $PbTiO_3$ by Lindenberg and co-workers in ultrafast time resolved optical-pump X-ray scattering-probe experiments.[47] In related work, long-lived charge carriers are known to trap to planar domain boundaries in ferroelectrics.[48–51] The net effect of such localization of photo-generated electrons and holes is to provide screening of the Coulomb potential. In line with the "ferroelectric polaron" proposal put forth by Miyata and Zhu,[7] this picture of dynamic charge carrier screening due to polar lattice domains or local ferroelectric domain boundaries may therefore underly the long-lived charge carrier dynamics observed in SbSI.

**Conclusion**

We have studied electron-soft phonon coupling and charge carrier dynamics in a ferroelectric semiconductor SbSI. By comparing polarization and probe-photon energy resolved coherent phonon spectra with corresponding Raman spectra, we provide direct, mode-specific experimental evidence for electron-soft phonon coupling. Furthermore, we discover exceptionally long-lived charge carrier dynamics associated with a transient electro-optical effect, which can be fully explained based on charge carrier interactions with thermally stimulated soft-phonon excitations. These findings provide insights into the photophysics of ferroelectric semiconductors and support the proposal that electron-soft phonon interactions can be harnessed to provide efficient charge carrier screening, leading to long charge carrier lifetimes in semiconductors.[7]

**Methods**

*Coherent Phonon/Transient Reflectance Spectroscopy*

The fundamental output of an ultrafast Ti:Sapphire regenerative amplifier (Coherent Legend 10 kHz, 5 mJ, seeded by Coherent Vitara oscillator) is split into two paths. One path of the fundamental is frequency doubled in a BBO crystal to generate a ~200 fs pump pulse centered at 3.1 eV, while several μJ from the other path is focused into a sapphire crystal to generate a white light continuum used for the probe pulse (see Supporting Information for time domain characterization of the pump and probe pulses). The polarization of the incident pump and probe pulses are controlled using half-waveplates, with an additional linear polarizer used for the probe. Pump and probe pulses are overlapped at near normal incidence on a SbSI crystal mounted in a continuous flow $LN_2$ cryostat (Oxford MicroStat Hi-Res). Reflected probe light is sent through a linear polarizer with polarization axis set to match the polarization of the incoming probe, and is then dispersed by a 600g/mm diffraction grating blazed at 500 nm and imaged onto a line-scan CCD camera (e2v AViiVA EM4). The pump pulse is chopped at half the laser repetition rate (5 kHz), and shot-to-shot synchronization with the line-scan camera is controlled by triggering a PCI-e framegrabber (NI PCIe-1433). The pump-probe delay is controlled using a mechanical delay stage in the pump line. The instrumentation is controlled and data recorded with custom written LabVIEW code, from which the pump-on vs. pump-off signal of the line-scan camera is used to calculate $\Delta R/R$.

*Polarization Resolved Raman Spectroscopy*

A 633 nm HeNe laser was used to measure low-frequency Raman scattering on a homemade set-up built around a Nikon TE-300 inverted microscope. A half-waveplate followed by a linear polarizer fixes the polarization at the HeNe laser source, which is sent to a reflective bandpass filter (Ondax), through a half-waveplate mounted on a motorized rotation stage used to control the angle of incident polarization in the sample plane, and into the microscope objective (40X, NA = 0.6). The objective is focused onto an SbSI crystal (the same used for coherent phonon spectroscopy) mounted in a continuous flow $LN_2$ cryostat (Oxford MicroStat Hi-Res). The backscattered light from the sample is collected through the same objective and sent back through the same motorized half-waveplate, which projects the parallel polarized Raman signal onto the polarization axis fixed at the HeNe laser, and the cross polarized Raman signal onto the orthogonal axis. The Raman signal is then sent through two sets of volumetric Bragg filters (Ondax) to filter out the Rayleigh scattered laser line and the parallel polarized Raman signal is selected using a half waveplate and fixed linear polarizer downstream of the Bragg filters. The Raman signal is collimated onto the entrance slit of a spectrometer (Princeton Instruments HRS-300) with a high-resolution holographic grating that disperses the spectrum onto a $LN_2$ cooled CCD camera (Princeton Instruments LN400/B).

*SbSI Synthesis and Characterization*

SbSI crystals were synthesized according to a previously published flux reaction[52] with 40 mol% of $Sb_2S_3$ in $SbI_3$. The powders were grinded, pressed into a pellet, and placed in a quartz tube. The pellet was melted with a propane torch under ~1 atm of nitrogen, cooled, and then melted again. Quartz wool was placed approximately halfway down the quartz tube and indents were made above the placement of the wool. The tube was sealed under partial pressure of nitrogen and placed in a box furnace. The reaction was heated to 475 °C over the course of a few hours, left at 475 °C for 2 hours, and ramped down to 350 °C at a rate of 0.5 °C/hr. The tube was taken out of the oven while at 350 °C and flipped over to filter the excess starting reagents with the use of the quartz wool. The crystals were washed with boiling methanol to remove leftover $SbI_3$. The SbSI crystals were characterized using an Agilent SuperNova single crystal x-ray diffractometer and compared to literature.

**Acknowledgements**

This work was supported by the Vannevar Bush Faculty Fellowship through the Office of Naval Research through Grant No. N00014-18-1-2080. Synthesis of the SbSI single crystal was supported by the Materials Science and Engineering Research Center (MRSEC) through NSF grant DMR-2011738. L.H. acknowledges support from the Swiss National Science Foundation under project ID 187996. M.E.Z. acknowledges postdoctoral fellowship support from the U.S. Department of Energy Office of Energy Efficiency and Renewable Energy administered by the Oak Ridge Institute for Science and Education (ORISE) for the DOE. ORISE is managed by Oak

Ridge Associated Universities (ORAU) under DOE Contract No. DE-SC0014664. All opinions expressed in this paper are the authors' and do not necessarily reflect the policies and views of DOE, ORAU, or ORISE. V.A.P. is supported by the National Science Foundation Graduate Research Fellowship Program (NSF GRFP #2019279091). J.C.R. is supported by the US Department of Defense through the National Defense Science & Engineering Graduate Fellowship (NDSEG) Program.

# Supporting Information

**Charge carrier coupling to the soft phonon mode in a ferroelectric semiconductor**

Mark E. Ziffer,[1] Lucas Huber,[1] Feifan Wang,[1] Victoria A. Posey,[1] Jake C. Russell,[1] Taketo Handa,[1] Xavier Roy,[1] and X.-Y. Zhu[1*]

1. Department of Chemistry, Columbia University, New York, NY 10027, USA
*To whom correspondence should be addressed: xz2324@columbia.edu

## Section S1: Softening of Coherent Phonons at 270K

In the lattice dynamical theory of soft-mode driven phase transitions, the soft phonon is modelled as a damped harmonic oscillator with a temperature dependent complex damping term that represents anharmonic phonon-phonon interactions which cause the phonon "softening".[1] As a result, towards the critical temperature ($T_c$) the soft-mode resonance frequency shifts and its dephasing time becomes shorter, both of which can be studied directly in the time domain from the dynamics in the coherent phonon spectrum at temperatures near $T_c$.

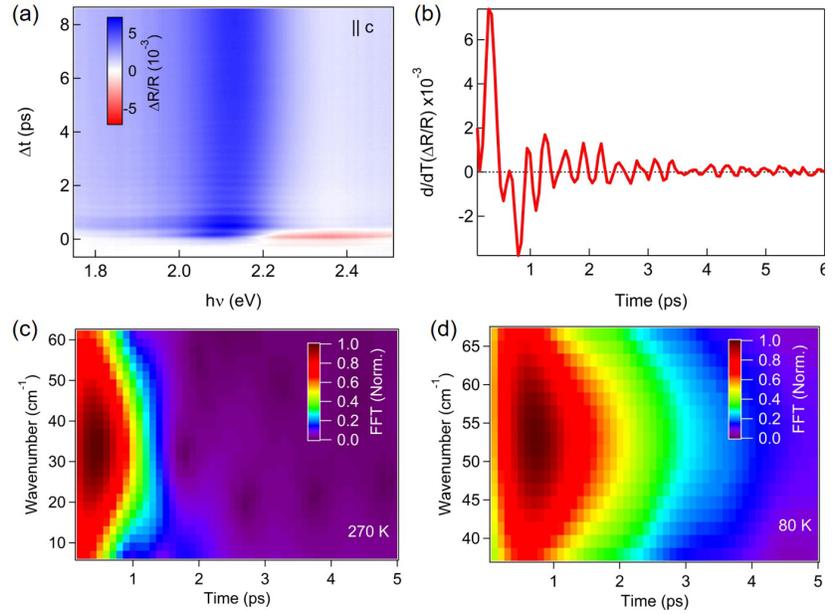

**Figure S1 (a.)** Transient reflectivity of SbSI measured at 270 K in the probe-$\parallel c$ polarization geometry. **(b.)** Time differentiated transient reflectance signal ($\frac{d}{dt}\Delta R/R$) from the data in **(a.)** spectrally integrated between 1.7-1.9 eV. **(c.)** STFT analysis of the 270K data in **(b.)**, with results cropped in the range of 7-65 cm$^{-1}$ corresponding to the STFT of the coherent soft-mode. **(d.)** STFT analysis of the below-gap 80 K coherent phonon data for the probe-$\parallel c$ polarization geometry (see main text Figure 2a.), with results cropped in the frequency range corresponding to the coherent soft-mode.

We characterize the softening behavior in the time domain by studying the coherent phonon response for the soft-mode at 270 K in the probe-$\parallel c$ polarization geometry. Figure S1a shows the full transient reflectivity data at 270 K in the probe-$\parallel c$ polarization geometry. Figure S1b shows the time differentiated transient reflectivity spectrally integrated between 1.7-1.9 eV, showing oscillations from the below-gap coherent phonon response. In Figure S1b., long period oscillations corresponding to the coherent soft mode are only evident before the first ~1.2 ps, while after ~1.2 ps only ~0.3 ps period oscillations corresponding to the ~115 cm$^{-1}$ coherent phonon can be observed. To fully analyze the dephasing dynamics of the coherent soft mode, we performed a short-time Fourier transform (STFT) analysis of the data in Figure S1b. Figure S1c. shows the results of the STFT cropped at the frequency range of the soft-phonon mode, which clearly shows a 1/e dephasing lifetime of ~1.2 ps corresponding to a resonance linewidth of ~28 cm$^{-1}$, which is in good agreement with the linewidth of the Raman spectrum at 270 K (~28 cm$^{-1}$) shown in Figure 1e of the main text. Furthermore, In Figure S1d we show the results of a STFT analysis of the coherent soft-phonon data in the probe-$\|c$ polarization geometry at 80 K (from the spectrally integrated below-gap data in Figure 2a. of the main text). Figure S1d shows that at 80 K, the soft-mode 1/e dephasing lifetime is ~2.9 ps corresponding to a linewidth of ~11 cm$^{-1}$, again in reasonably good agreement with the linewidth of the soft-phonon Raman spectrum at 77 K (~6 cm$^{-1}$). We note that the choice of window length in STFT analysis can affect the results and could be responsible for the small differences in linewidth calculated from the STFT versus those measured in Raman spectroscopy; however we found that the results in the time domain converge for a window length of up to 2 ps used for the analysis in Figures S1c and d.

## Section S2: 80K Transient Reflectance Spectrum

In Figure S2 we show the full transient reflectivity data at 80 K in the probe-$\perp c$ polarization geometry which was used for comparison with the electroreflectance spectrum in Figure 4d of the main text. The spectrum shown in Figure 4d of the main text is spectrally integrated between 0.5-3 ns. We note that there is a significant dynamic redshift of the transient reflectivity spectrum at pump-probe delay times between 0-0.5 ns. This redshift could be due to several effects, such as slow lattice cooling or potentially long-lived hot-electrons, and are subject to future study.

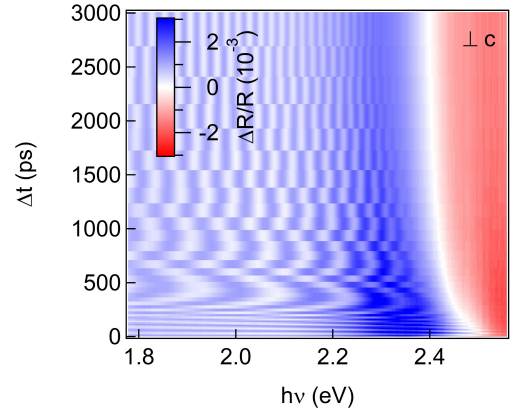

**Figure S2.** Transient reflectance data from SbSI at 80K for probe polarized perpendicular to the *c*-axis. Color bar represents intensity of $\Delta R/R$.

## Section S3: Time Domain Pulse Characterization

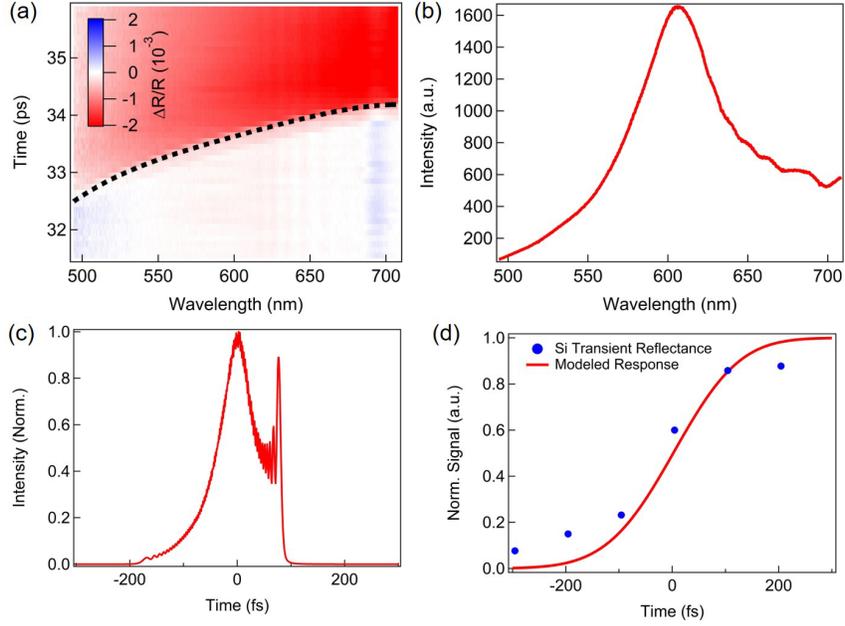

**Figure S3 (*a.*)** Raw transient reflectivity data on a silicon wafer (native oxide), with the signal onset time as a function of probe wavelength indicated by the dashed black line. Color bar represents intensity of $\Delta R/R$. (***b.***) Spectrum of the white light probe pulse taken from the transient reflectance measurement on silicon. (***c.***) Reconstructed time domain probe pulse intensity profile calculated from the probe intensity spectrum and spectral phase extracted from (*a.*) and (*b.*). (***d.***) Spectrally integrated transient reflectance signal rise from Si compared with the convolution of the pump-probe cross correlation with a Heaviside step function for a Gaussian pump pulse profile with 200 fs FWHM.

We used the method of Wegkamp *et al.*[2] to first measure the spectral phase of the white light pulse by evaluating the probe-wavelength dependent onset of the transient reflectance signal rise time from a silicon substrate. The transient reflectance signal onset for each probe wavelength is defined by the maximum of the first derivative of the transient reflectance signal with respect to time. The signal onset versus time gives the frequency dependent group delay of the probe pulse $\tau_g(\nu)$ as in Figure S3a, which is used to calculate the spectral phase $\varphi(\nu) = \int \tau_g(\nu) d\nu$, where $\nu = 2\pi f$.[2] The white light pulse spectrum taken from the raw transient reflectance probe intensity (Figure S3b) is then combined with the spectral phase to give the complex pulse profile in the frequency domain, which is used to retrieve the probe pulse intensity profile in the time domain (Figure S3c) via the inverse Fourier transform $I(t) \propto \left|\int_0^\infty \sqrt{I(\nu)} e^{i\varphi(\nu)} e^{-2\pi i \nu t} d\nu\right|^2$. To determine the pump pulse intensity profile, we take the cross correlation of the probe pulse profile with a Gaussian pulse function that represents the pump pulse profile, and convolve this modeled pump-probe cross correlation with a Heaviside step function which represents the ultrafast electronic response in silicon.[3] We then vary the parameters of the Gaussian pulse function and compare the results with the spectrally integrated transient reflectance rise time for silicon (taken from the data in Fig S3a). Figure S3d shows the modeled response assuming a pump pulse profile with a Gaussian width of 200 fs (FWHM), which gives good agreement with the silicon transient reflectance rise time.

**Supplementary References:**